\documentclass{article}
\usepackage[utf8]{inputenc}
\usepackage{amsmath}
\usepackage{color}
\usepackage{mathrsfs}
\usepackage{amsmath}
\usepackage{hyperref}
\usepackage{bigints}
\DeclareMathOperator\arctanh{arctanh}
\usepackage{amssymb}
\usepackage[top=2cm, bottom=2.5cm, left=2cm, right=2cm]{geometry}
\usepackage{graphicx}
\usepackage{mwe}
\usepackage{amsbsy}
\usepackage{graphicx}  
\usepackage{float}
\usepackage{comment}
\usepackage{caption}
\usepackage[usenames,dvipsnames]{pstricks}
\usepackage{pstricks-add}
\usepackage{epsfig}
\usepackage{pst-grad} 
\usepackage{pst-plot} 
\usepackage[space]{grffile} 
\usepackage{etoolbox} 
\makeatletter 
\patchcmd\Gread@eps{\@inputcheck#1 }{\@inputcheck"#1"\relax}{}{}
\makeatother
\usepackage[usenames,dvipsnames]{pstricks}
\usepackage{pstricks-add}
\usepackage{epsfig}
\usepackage{pst-grad} 
\usepackage{pst-plot} 
\usepackage[space]{grffile} 
\usepackage{etoolbox} 
\makeatletter 
\patchcmd\Gread@eps{\@inputcheck#1 }{\@inputcheck"#1"\relax}{}{}
\makeatother

\title{Transparent Spheres as Gravitational Lens}

\author{Edwin Santiago-Leandro, Alexander Mora-Chaverri, Francisco Frutos-Alfaro}

\date{\today}

\begin{document}

\maketitle


\begin{abstract}
In this contribution, we present a short account of gravitational lenses and how to calculate different properties of its images in the case of having a transparent distribution of matter such as the uniform transparent sphere, isothermal gas sphere, non-singular isothermal gas sphere and a transparent King profile. With the help of XFGLenses software, and numerical methods, different images arising from all of these profiles, and the different caustics and critical curves are shown. The images were consistent with several previous results that are expected for transparent profiles, like having an odd number of images, and reducing the number of images by two when the source passes through the caustic. The curves shown in the caustics where the diamond, the ellipse and the lemniscate-like. For the critical curves, the most common curve was the ellipse, and the lemniscate appeared in the transparent NSIS case, which is consistent with the fact that these curves are common in gravitational lenses.

\end{abstract}

\maketitle

\section{Introduction}

The treatment of gravitational lenses as transparent objects has not been considered as much as the opaque case. In 1971, Clark considered the uniform and transparent gravitational lens and calculated the deflection angle and the effects on beam area, apparent luminosity and focusing effects \cite{Clark}. Bourassa and Kantowski made further calculations on cases with spheroidal symmetry \cite{BK1, BK2}. After this in 1984, there was a minor correction made by Bray in the surface density integral made by Bourassa and Kantowski \cite{Bray}. Another article regarding the subject was made by McKenzie \cite{McKenzie}, where he shows that transparent lenses make an odd number of images. This was also shown by Dyer and Roeder in 1980 \cite{DR}. Nandor and Helliwell analyzed gravitational lensing with Fermat's principle and they used the model of a transparent lens with a logarithmic varying thickness \cite{NH}.  

There have also been interesting experiments where people reproduce images with transparent objects, simulating the effects realistically. In 1969, Liebes, used Plexiglas to simulate magnification and lensing effects like the ones he studied previously in 1964 \cite{L1,L2}. Icke constructed a cilyndrical lens, also employing Plexiglas, to approximate the lensing of a point mass object in 1980 \cite{Icke}. Higbie used Plexiglass too \cite{Hig1}, and Falbo-Kenkel and Lohre used bases of wine glass to simulate gravitational lensing \cite{FKL}. Adler et al. employed plastic lenses according to their calculations for the point mass, the constant density sphere and the isothermal gas sphere \cite{ABR}. Recently, Selmke used a setup with a water filled acrylic pool and small discs to replicate the case of a single, a binary and a triple mass lens \cite{Selmke}. Having said that, there have not been much done regarding computer simulations of these images caused by the transparent cases.

Adler et al. calculated the deflection angle for the transparent uniform sphere, the transparent singular isothermal gas sphere (SIS), and the non-singular case (NSIS) \cite{ABR}. In this article, using the results of Adler et al., we develop and get the calcultations that are necessary to plot the caustics and critical curves. The transparent King profile, and its derivatives are also included. From these results, images are obtained for the different transparent profiles, with some caustics and critical curves associated with these profiles. For the computational images, the simulator XFGLenses \cite{Frutos} is employed and for some of the critical curves and caustics, a code made in MATLAB is used to find the curves. XGFLenses is a computational tool capable of producing the images that produces a certain galaxy or galaxy cluster given the density distribution and the parameters relevant for the calculation of the images. It is also capable of calculating the caustics of a given gravitational lens. Our aim is to make computer simulated images made with XGFLenses of different transparent cases, including the transparent and uniform sphere, and the non uniform cases with spheroidal symmetry. In Section 2, a brief description of the gravitational lens theory is given. In section 3, the transparent uniform sphere is discussed. In following sections, the results for the mentioned lens models are presented. In Section 7, we will show the images and analyze them, and in Section 8, some conclusions are given.

\section{Gravitational lenses}

\subsection{Deflection angle}

For the case of a point mass, using the Schwarzschild metric, it can be shown from the geodesic equation that the deflection angle is
given by

\begin{equation}
    \hat{\alpha} = \frac{4GM}{c^{2}R},
\end{equation}

\noindent
where $M$ is the mass of the deflecting object and $R$ is the closest distance from the light to the object (impact parameter). From a mass distribution, it is possible to generalize this result by making the following integration 

\begin{equation}
      \hat{\boldsymbol{\alpha}}(\boldsymbol{\xi})= \frac{4G}{c^{2}} \int_{\mathbb{R}^{2}} \Sigma(\boldsymbol{\xi}') \frac{(\boldsymbol{\xi}-\boldsymbol{\xi}')}{\arrowvert \boldsymbol{\xi}-\boldsymbol{\xi}' \arrowvert^{2}}d^{2}\xi',
\end{equation}

\noindent
where $\Sigma( \boldsymbol{\xi}')$ is the surface mass distribution. Even though it is clear that the mass distribution is a volumetric distribution, we can use the fact that the size of the lens is very small compared to the cosmological distances from the lens to the source plane and to the observer. This is known as the {\it{thin lens approximation}}.

\noindent
In this article, we calculate different deflection angles but in all cases the matter distributions are axially symmetric. This allows the deflection angle equation to be reduced to 

\begin{equation}
    {\hat{\alpha}}(r) = \frac{4GM(r)}{c^2r},
\end{equation}

\noindent
where $M(r)$ is the projected mass that is enclosed a distance $r$ from the origin. By calculating $M(r)$ in this cases, the deflection angles are determined. Actually, there are actually two ways of calculating $M(r)$. For a tridimensional mass distribution $\mu(r)$, one can calculate the projected bidimensional distribution $\Sigma(\xi)$ applying an Abel Transformation \cite{Karttunen} to find $M(r)$, or, alternatively, one can perform the tridimensional integration with $\mu(r)$, making sure that your integration only encompasses the enclosed mass in a cylinder of radius $r$ as seen from Earth (see Figure 1).

\begin{figure}[H]
    \centering
    \includegraphics[width=10cm]{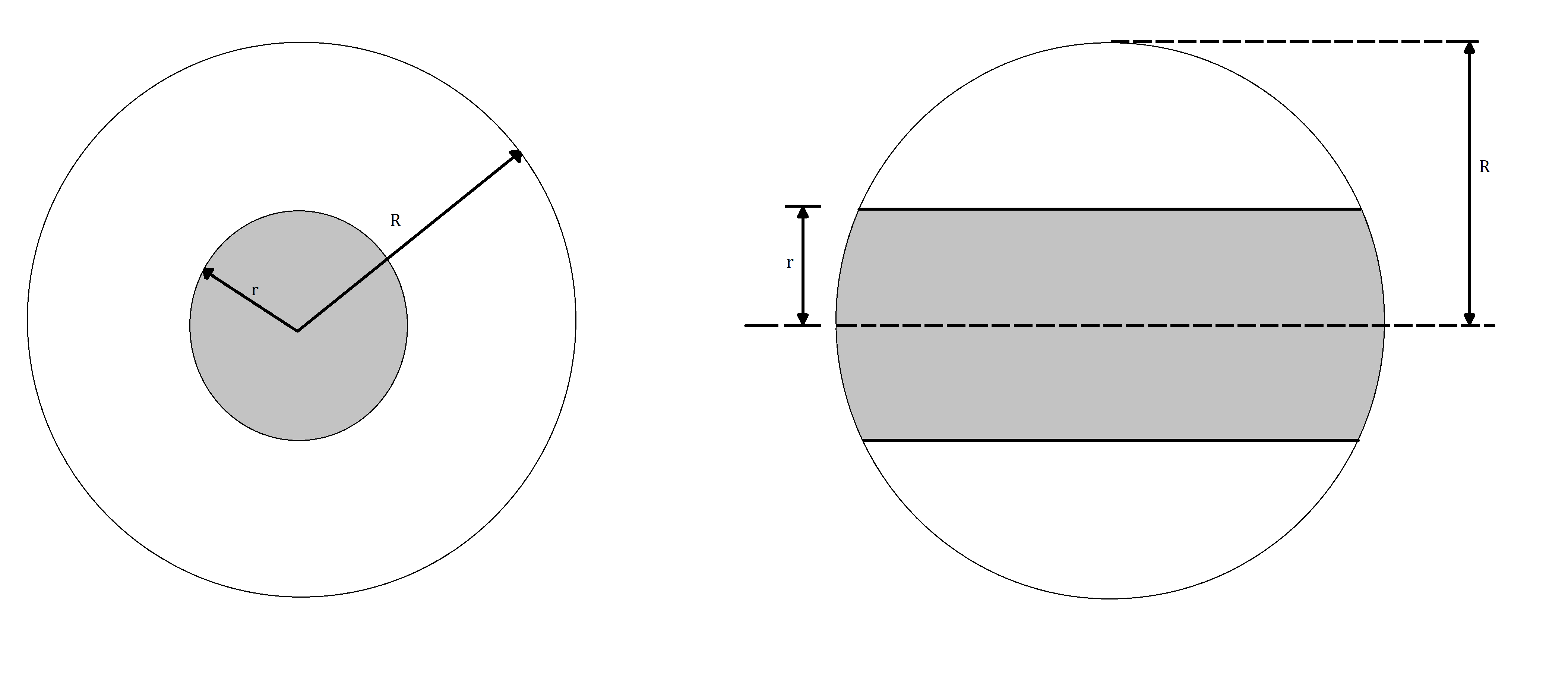}
    \caption{Diagram of the enclosed mass in a cylinder of radius $r$. On the left, you can see the projected mass, as seen from Earth. On the right, you can see the lateral view.}
    \label{fig:my_label}
\end{figure}

\subsection{Lens equation}

\noindent
Although gravitational lensing is an effect from general relativity, it is possible to make a purely geometric relation. Using the notation from Figure 2, the following geometric relation can be written

\begin{equation}
\label{lenseq1}
  \boldsymbol{\beta}=\boldsymbol{\theta} 
  - \frac{D_{LS}}{D_{S}} \hat{\boldsymbol{\alpha}} . 
\end{equation}

\noindent
From (\ref{lenseq1}), the dimensionless lens equation is obtained

\begin{equation}
\label{lenseq2}
  \boldsymbol{y}=\boldsymbol{x}- \boldsymbol{\alpha}(\boldsymbol{x}),
\end{equation}

\begin{figure}[H]
    \centering
    \includegraphics[width=8cm]{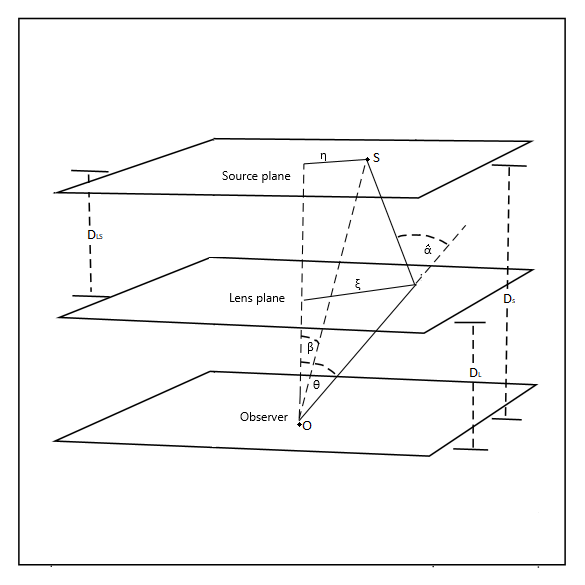}
    \caption{Gravitational deflection angle}
\end{figure}

\noindent
where 

$$ \boldsymbol{\alpha}(\boldsymbol{x}) = \frac{D_{L}D_{LS}}{\xi_{0}D_{S}} {\hat{\boldsymbol{\alpha}}(\xi_{0}\boldsymbol{x})} =  \int_{\mathbb{R}^{2}} \kappa(\boldsymbol{x}') \frac{(\boldsymbol{x}-\boldsymbol{x}')}{\arrowvert \boldsymbol{x}-\boldsymbol{x}' \arrowvert^{2}}d^{2} x', $$ 

$$ \boldsymbol{y}\equiv \frac{\boldsymbol{\eta}}{\eta_{0}} \qquad {\rm and} \qquad \boldsymbol{x}\equiv \frac{\boldsymbol{\xi}}{\xi_{0}} . $$ 

\noindent
The parameter $\xi_0 $ is the length scale on the lens surface. The dimensionless surface density of the lens is given by

\begin{equation}
\kappa(\boldsymbol{x}) = \frac{\Sigma(\boldsymbol{\xi})}{\Sigma_c},
\end{equation}

\noindent
where $\Sigma_{c}$ represents the critical density, which is given by

\begin{equation}
{\Sigma_c} = \frac{c^2 D_S}{4 \pi G  D_L D_{LS}}.
\end{equation}

\noindent
A generalization of the lens equation (\ref{lenseq2}) is the following

\begin{equation}
\label{gleq}
      {\boldsymbol{y}} = {\boldsymbol{M}} \cdot {\boldsymbol{x}} - {\boldsymbol{\alpha}}({\boldsymbol{x}}) ,
\end{equation}

\noindent
where $ {\boldsymbol{x}} = (x_1, \, x_2) $ is the image position on the lens plane
($x^2 = x_1^2 + x_2^2$), $ {\boldsymbol{y}} = (y_1, \, y_2) $ is the source position
on the source plane, and the matrix $ {\boldsymbol{M}} $ is given by

\begin{equation}
\label{matrix}
{\boldsymbol{M}} = \left(
\begin{array}{cc}
1 - \sigma - \gamma \cos{2 \phi} &            - \gamma \sin{2 \phi} \\
           - \gamma \sin{2 \phi} & 1 - \sigma + \gamma \cos{2 \phi}
\end{array}
\right)  
,
\end{equation}

\noindent
where $ \sigma $ is the dimensionless macrolens density, $ \gamma $ is the dimensionless ma\-cro\-lens shear, and $ \phi $ is the shear angle. The components $M_{ij}$ ($i,j=(1,2)$) of the matrix are 

\begin{eqnarray}
\label{compm}
M_{11} & = & 1 - \sigma - \gamma \cos{2 \phi}, \nonumber \\
M_{12} & = & M_{21} = - \gamma \sin{2 \phi}, \\
M_{22} & = &1 - \sigma + \gamma \cos{2 \phi} . \nonumber
\end{eqnarray}

\subsection{Caustics and critical curves}

The critical and caustic curves are important in gravitational lensing analysis, because they divide the lens into regions of interest. Knowing the caustics and the position of the source, the number of images that a gravitational lens will create can be determined, as well as where the images will be magnified \cite{SEF}.

The critical curves are the curves that are formed in the lens plane, on the other hand, the caustics are formed in the source plane \cite{SEF}. The critical curves can obtained if the determinant of the Jacobian is equal to zero, that is $ {\rm det}(\boldsymbol{J}) = 0$ and the matrix $\boldsymbol{J}$ is given by

\begin{equation*}
\boldsymbol{J}=
\begin{pmatrix}
J_{11} & J_{12}\\
J_{21} & J_{22}
\end{pmatrix}=\frac{\partial \boldsymbol{y}}{\partial \boldsymbol{x}}.
\end{equation*}

\noindent 
The components of the Jacobian $ {\boldsymbol{J}} $ are

\begin{eqnarray}
\label{compj}
J_{11} = \frac{\partial y_1}{\partial x_1} & = &
\left(M_{11} - \frac{\alpha}{x} \right)
- \frac{x_1^2}{x^2} \left(\frac{d \alpha}{d x} - \frac{\alpha}{x} \right) ,
\nonumber \\
J_{12} = \frac{\partial y_1}{\partial x_2} & = & M_{12}
- \frac{x_1 x_2}{x^2} \left(\frac{d \alpha}{d x} - \frac{\alpha}{x} \right) ,
\nonumber \\
J_{21} = \frac{\partial y_2}{\partial x_1} & = & M_{21}
- \frac{x_1 x_2}{x^2} \left(\frac{d \alpha}{d x} - \frac{\alpha}{x} \right) ,
\\
J_{22} = \frac{\partial y_2}{\partial x_2} & = &
\left(M_{22} - \frac{\alpha}{x} \right)
- \frac{x_2^2}{x^2} \left(\frac{d \alpha}{d x} - \frac{\alpha}{x} \right) .
\nonumber
\end{eqnarray}

\noindent
The determinant of the Jacobian is

\begin{eqnarray}
\label{detjac}
{\cal J} & = & {\rm{det}}  \hspace{0.1cm}{\boldsymbol{J}} = J_{11} J_{22} - J_{12} J_{21} \\
& = & \left[\left(M_{11} - x_2^2 \frac{\alpha}{x^3} \right)
- \frac{x_1^2}{x^2} \frac{d \alpha}{d x} \right] \nonumber \\
& \times & \left[\left(M_{22} - x_1^2 \frac{\alpha}{x^3} \right)
- \frac{x_2^2}{x^2} \frac{d \alpha}{d x} \right] \nonumber \\
& - & \left[M_{12} - \frac{x_1 x_2}{x^2} \left(\frac{d \alpha}{d x}
- \frac{\alpha}{x} \right) \right]  \nonumber \\
& \times & \left[M_{12} - \frac{x_1 x_2}{x^2} \left(\frac{d \alpha}{d x}
- \frac{\alpha}{x} \right) \right]. \nonumber
\end{eqnarray}

\noindent
The caustics are mapped using the lens equation, in which the $x_{1}$ and $x_{2}$ are evaluated, canceling the determinant of the Jacobian matrix. Caustic curves are generally described by polar curves, therefore using a change of variable of $x_{1}$ and $x_{2}$ in terms of $\theta$  and $r'$,

\begin{eqnarray}
    x_{1} &=& r' \cos{\theta} \nonumber , \\
    x_{2} &=& r' \sin{\theta} .
\end{eqnarray}

\noindent
From (\ref{gleq}), (\ref{compm}) and (\ref{compj}), we can obtain a general expression for the determinant given by 

\begin{equation}
\label{detj}
      \text{det} \, \boldsymbol{J} = \text{det} \, \boldsymbol{M} + \left(\frac{d\alpha}{dr'}- \text{Tr} \, \boldsymbol{M} \right) \frac{\alpha}{r'}-\left(\frac{d\alpha}{dr'}-\frac{\alpha}{r'}\Bigg)\Bigg(\frac{1}{2} \text{Tr} \, \boldsymbol{M}+ \gamma \cos(2(\theta-\phi)) \right) .
\end{equation}

\noindent
Considering the case in which $\boldsymbol{M}$ is equal to identity matrix, equation (\ref{detj}) does not depend on $\theta$, reducing to the following expression

\begin{equation}
     \text{det} \, \boldsymbol{J} = \left(\frac{\alpha}{r'}-1 \right) \left(\frac{d\alpha}{d r'}-1 \right) .
\end{equation}

\noindent
In this case, the form of the caustics can be obtained from

\begin{equation}
    \left(\frac{d\alpha}{d r'}-1 \right) \left(\frac{\alpha}{r'}-1 \right) = 0 .
\end{equation}

\section{Transparent Sphere}

As first example, the transparent sphere is considered. Whether it is the transparent or opaque case, a uniform matter distribution is useful as a first case to study, because it is among the easiest mass distributions. 

\noindent
The transparent sphere density is given by

\begin{equation}
\label{rho}
\mu(r) = \left \{
\begin{array}{ll}
M/V & r \le R \\
0   & r > R \\
\end{array}
\right.,
\end{equation}

\noindent
where $ M $ is its mass, $ R $ its radius, and $ V = 4 \pi R^3 / 3 $ its volume. For simplicity, we will define 

$$\mu=M/V.$$

\noindent
Because the transparent uniform sphere is axially symmetric, we only need to pay attention to the enclosed mass a distance $r$ from the center of the distribution, so it can be scaled to find the deflection angle as a function of the scaled radius $x$.

\noindent
The integration for $M(r)$ gives

\begin{equation}
    M(r)= 4\pi \int_{0}^{r}\int_{0}^{\sqrt{R^2-\rho^2}} \mu \rho dz d\rho = M\left[1 - \left(1-\frac{r^2}{R^2}\right)^{3/2}  \right].
\end{equation}

\noindent
Then, the deflection angle would be

\begin{equation}
    \alpha=  \frac{2R_{s}}{r}\left[1 - \left(1-\frac{r^2}{R^2}\right) ^{3/2} \right].
\end{equation}

\noindent
The final step is to scale in terms of $x$. The way this is done is by defining $x=r/(2R_{s})$, and $x_{0} = R/(2R_{s})$. From this, the Einstein angle becomes 

\begin{equation}
\label{alpha}
\alpha(x) = \left \{
\begin{array}{ll}
\displaystyle{\frac{1}{x}} \left[1 - \left(1 - \frac{x^2}{ x_0^2} \right)^{3/2} \right]  & x \le x_0, \\
\displaystyle{\frac{1}{x}} & x > x_0, \\
\end{array}
\right.
\end{equation}

\noindent
where $ x, \, x_0 $ are the new scaled variable and the new scaled radius, respectively. The derivative is

\begin{equation}
\label{dalpha}
\frac{d \alpha}{d x} =
\left \{
\begin{array}{ll}
- f(x) & x \le x_0, \\
- \displaystyle{\frac{1}{x^2}} & x > x_0, \\
\end{array}
\right.
\end{equation}

\noindent
with

\begin{equation}
\label{func}
f(x) = \frac{1}{x^2}
\left[1 - \left(1 - \frac{x^2}{x_0^2} \right)^{3/2} \right]
- \frac{3}{x_0^2} \left(1 - \frac{x^2}{x_0^2} \right)^{1/2} .
\end{equation}

\section{Transparent Isothermal Gas Sphere }

The isothermal gas sphere or SIS describes a relatively simple distribution of matter with certain realistic properties. It is an axially symmetric distribution that gives flat rotation curves. This is important to describe the dark matter halo in galaxies. The name stems from the fact that it also represents a distribution of gas where the pressure is proportional to its density \cite{ABR, SEF}.

\noindent
The density profile for the SIS is

\begin{align}
    \mu(r) &= \frac{b}{r^2} = \frac{b}{\rho^{2}+z^2},
\end{align}

\noindent
where $r^2=\rho^2+z^2$. Here, the deflection angle becomes \cite{ABR}

\begin{equation}
    \hat{\alpha} (r) = \frac{2R_{s}}{r}\left(1-\left(1-\frac{r^2}{R^2} \right)^{\frac{1}{2}}+ \frac{r}{R} \arccos\left(\frac{r}{R}\right) \right),
\end{equation}

\noindent
with the total mass is $M=4\pi b R$. Now, the scaled deflection angle is

\begin{equation}
    \alpha(x) = \frac{1}{x}\left(1-\left(1-\frac{x^2}{x_{0}^2} \right)^{\frac{1}{2}}+ \frac{x}{x_{0}} \arccos\left(\frac{x}{x_{0}}\right) \right).
\end{equation}

\noindent
The derivative of the deflection angle is

\begin{equation}
    \frac{d\alpha}{dx}=\frac{-1}{x^2}\left(1-\left(1-\frac{x^2}{x_{0}^2}\right)^{\frac{1}{2}} \right).
\end{equation}

\section{Transparent non Singular Isothermal Gas Sphere}

The NSIS generalizes the SIS by adding a core radius, and it is a more realistic model for a mass distribution. Moreover, there is a disadvantage, since the SIS density diverges at $r=0$. This is another reason why the NSIS distribution is studied. For the NSIS distribution, the density profile is given by

\begin{equation}
    \mu (\rho,z) = \frac{b}{\rho^{2}+z^2+r_{c}^2} ,
\end{equation}

\noindent
where ${r_{c}}$ is the core radius. The deflection angle is \cite{ABR} 

\begin{equation}
    \hat{\alpha} (r) = \frac{2R_{s}}{r \beta_{3}} \left( {1-\left(1-\frac{r^2}{R^2} \right)^{\frac{1}{2}}+\frac{\sqrt{r^2+r_{c}^2}}{R}\arccos{\beta_1}
    -\frac{r_{c}}{R}\arccos{\beta_2}}\right).
\end{equation}

\noindent
Scaling, the deflection angle becomes 

\begin{equation}
    \alpha (x) = \frac{1}{x \beta_{3}} \left( {1-\left(1-\frac{x^2}{x_{0}^2} \right)^{\frac{1}{2}}+\frac{\sqrt{x^2+x_{c}^2}}{x_{0}}\arccos{\beta_1}-\frac{x_{c}}{x_{0}}\arccos{\beta_2}}
    \right),
\end{equation}

\noindent
where  

\begin{eqnarray}
x_{c} &=&r_{c}/(2R_{s}) , \nonumber \\   
\beta_1 &=& \sqrt{\frac{x^2+x_{c}^2}{x_0^2+x_{c}^2}}, \nonumber \\ 
\beta_2 &=& \frac{x_{c}}{\sqrt{x_0^2+x_{c}^2}}, \\
\beta_{3} &=& 1-\frac{x_{c}}{x_{0}} \arccos{\beta_2}. \nonumber
\end{eqnarray}

Note that if $x_{c}=0$, the deflection angle is the SIS angle. Its derivative is

\begin{equation}
\frac{d\alpha}{dx}= \frac{1}{ x^{2}\beta_{3}} \left(
\frac{x^2}{x_{0} f_1} \arccos{\beta_1} + \frac{x^2}{x_{0}^2 f_0} + f_0 -\frac{x^2}{x_{0} f_2f_3}
-\frac{f_1}{x_{0}}\arccos{\beta_1}-\beta_{3} \right) ,
\end{equation}

\noindent
where

\begin{eqnarray}
    f_0 &=& \sqrt{1-\frac{x^2}{x_{0}^2}}, \nonumber \\ 
    f_1 &=& \sqrt{x^2+x_{c}^2}, \\ 
    f_2 &=& \sqrt{x_{0}^2+x_{c}^2}, \nonumber \\ 
    f_3 &=& \sqrt{1-\frac{x^2+x_{c}^2}{f_2^{2}}}. \nonumber
\end{eqnarray}

\section{Transparent King Model}

The King density model was proposed as a distribution that maps the density profile of the Coma cluster of galaxies \cite{King}. Moreover, it is advantageous, because it does not diverge at $r=0$, and it works well in general as a density model for clusters of galaxies with a flat rotation curve. The King density model is given by \cite{King}

\begin{equation}
    \mu (r') = \frac{\mu_{0}}{(r'^{2}+1)^{\frac{3}{2}}},
\end{equation}

\noindent
where $r'= {(\rho^{2}+z^{2})}/{r_{c}}$ is the scaled distance from the center and $\rho$ and $z$ are the cylindrical coordinates. 

\noindent
The deflection angle becomes

\begin{equation}
    \hat{\alpha} (r)=\frac{16\pi G\mu_{0}r_{c}^{3}}{c^{2}r}\left(\frac{1}{\sqrt{r_{c}^2+R^2}}\left(\sqrt{R^{2}-r^{2}}-R\right)\arctanh{\left(\frac{R}{\sqrt{r_{c}^2+R^2}}\right)}-\arctanh{ \left(\frac{\sqrt{R^{2}-r^{2}}}{\sqrt{r_{c}^2+R^2}}\right)}\right). 
\end{equation}

\noindent
By evaluating to find the total enclosed mass $M(r)$ in $r=R$ gives 

\begin{equation}
    M=4 \pi \mu_{0} r_{c}^3 \left(\arctanh{\left(\frac{R}{\sqrt{r_{c}^2+R^2}}\right)} -\frac{R}{\sqrt{r_{c}^2+R^2}} \right) = 4 \pi \mu_{0} r_{c}^3 f_4,
\end{equation}

\noindent
with this definition, we can write the deflection angle as

\begin{equation}
\hat{\alpha} (r) = \frac{2R_{s}}{r}\left[1+ \frac{1}{f_4} \sqrt{\frac{R^{2}-r^{2}}{r_{c}^2+R^2}} - \frac{1}{f_4} \arctanh{\left(\sqrt{\frac{R^{2}-r^{2}}{r_{c}^2+R^2}}\right)} \right].    
\end{equation}

\noindent
Scaling, and using the definitions $f_{i}$ defined for the NSIS case, we obtain

\begin{equation}
\alpha(x) = \frac{1}{x}\left[1+\frac{x_{0}f_{0}}{f_{5}}-\frac{f_{2}}{f_{5}} \arctanh\left(\frac{x_{0}f_{0}}{f_{2}}\right) \right],     
\end{equation}

\noindent
where

\begin{equation}
    f_{5}=f_{2}\arctanh\left(\frac{x_{0}}{f_{2}}\right)-x_{0} .
\end{equation}

\noindent
The derivative of the deflection angle is

\begin{equation}
    \frac{d\alpha}{dx}=-\frac{1}{x^2}\left( 1-\frac{f_{2}^2x^2}{x_{0}f_{0}f_{5}(f_{2}^{2}-x_{0}^{2}f_{0}^{2})}+\frac{x^2}{x_{0}f_{0}f_{5}}-\frac{f_{2}}{f_5} \arctanh{\left(\frac{x_{0}f_{0}}{f_{2}} \right)}+\frac{x_{0}f_{0}}{f_{5}}\right).
\end{equation}

\section{Images, Caustics and Critical Curves for the profiles}

Now, we proceed to analyze the images, caustics and critical curves generated by the profiles described above. Figures 3, 5, 7, 9 and 11 were generated with XFGLenses. For Figures 4, 6, 8, 10 and 12, MATLAB was employed, because the caustics and critical curves for the transparent version of these profiles are not implemented in XFGLenses yet. In Figure 3, a ring is formed if the source is projected at the center of the mass distribution with the parameters $\gamma$ and $\sigma$ are null. This is known in gravitational lens theory as an Einstein Ring  \footnote[1]{In fact, Chwolson was the first to publish that a ring could form if the lens, the source and the observer are in the same line of vision. \cite{Chw}} \cite{SEF}. Note that the first image in Figure 5 also has the source in the origin. The difference between this image and the one from Figure 3 is that a non-zero value for $\sigma$ and $\gamma$ is added. The images are still connected but one can still note that there are four distinct images surrounding the origin and a small image in the origin, having five images in total. This is expected because, as we mentioned in the introduction, a transparent lens should have an odd number of images \cite{McKenzie}.

The positions of the caustics of the uniform transparent sphere for $\boldsymbol{M}=\boldsymbol{I}$ (identity matrix) are calculated. Following from (\ref{detj}), and using (\ref{alpha}) and (\ref{dalpha}), we have two equations when $x<x_{0}$ that give possible critical curves:

\begin{align}
\label{eq1}
    1-\left(1-\frac{r'^2}{x^{2}_{0}} \right)^{\frac{3}{2}}-r'^{2}&=0 , \\
\label{eq2}    
    \frac{3}{x_{0}^{2}}\left( 1-\frac{r'^2}{x_{0}^{2}}\right)^{\frac{1}{2}}-\frac{1}{r'^{2}}\left[1-\left(1-\frac{r'^2}{x_{0}^{2}}\right)^{\frac{3}{2}} \right]-1 &=0 .
\end{align}

\noindent
The first equation has one solution for $1<x_{0}<\sqrt{\frac{3}{2}}$, and the second equation has one solution when $0<x_{0}<\sqrt{\frac{3}{2}}$. Now, let us consider the case $x>x_{0}$. In this case, we have the point mass case and the equations would be:

\begin{align}
\label{eq3}
    1-\frac{1}{r'^2} &= 0 , \\
\label{eq4}    
    1+\frac{1}{r'^2} &= 0 .
\end{align}

\noindent
In this case, it's obvious that (\ref{eq3}) has one solution $r'=1$ and (\ref{eq4}) has no solutions. The conclusion from this is that for $0<x_{0}<1$, there are two critical curves which are circles whose radii are the solutions of (\ref{eq2}) and (\ref{eq3}); and, for $1<x_{0}<\sqrt{\frac{3}{2}}$, there are also two critical curves, which correspond to circles whose radii are the solutions of (\ref{eq1}) and (\ref{eq2}). Figure 4 shows these results.

\begin{figure}[H]
\centering
    \includegraphics[width=.30\textwidth]{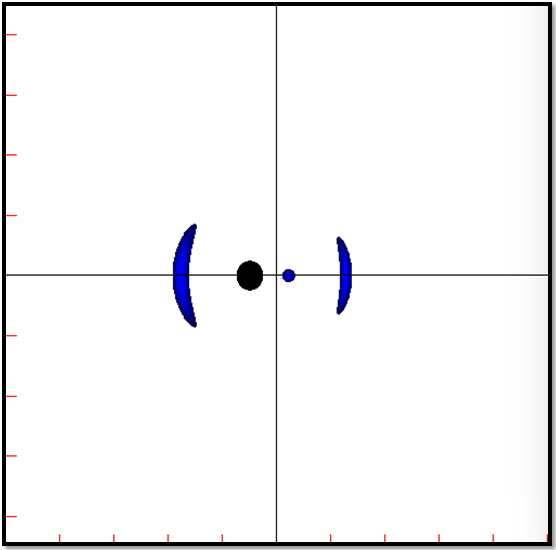}
    \includegraphics[width=.30\textwidth]{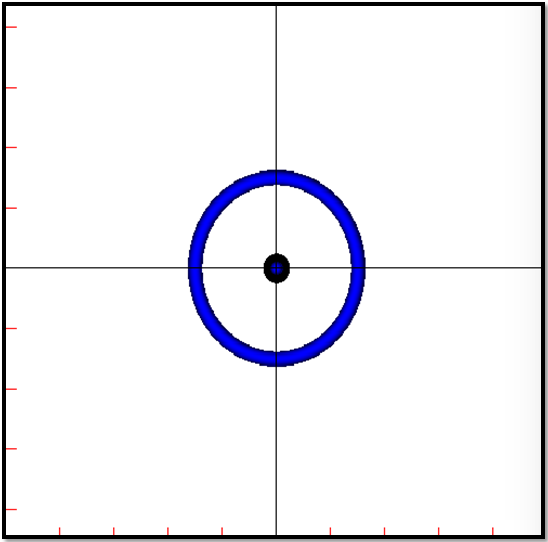}  
    \includegraphics[width=.30\textwidth]{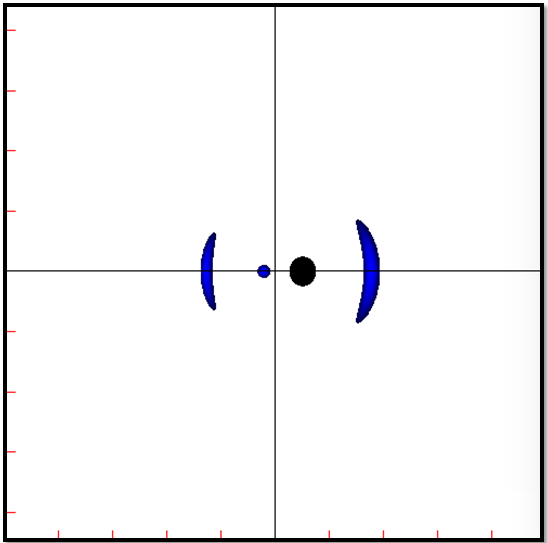}
    \caption{Images for the uniform transparent sphere. The black dot represents where the original source is with respect to the center of the distribution. The values for the parameters are null, which means that $\boldsymbol{M}=\boldsymbol{I}$.}
\end{figure}

\begin{figure}[H]
\centering
    \includegraphics[height=5cm]{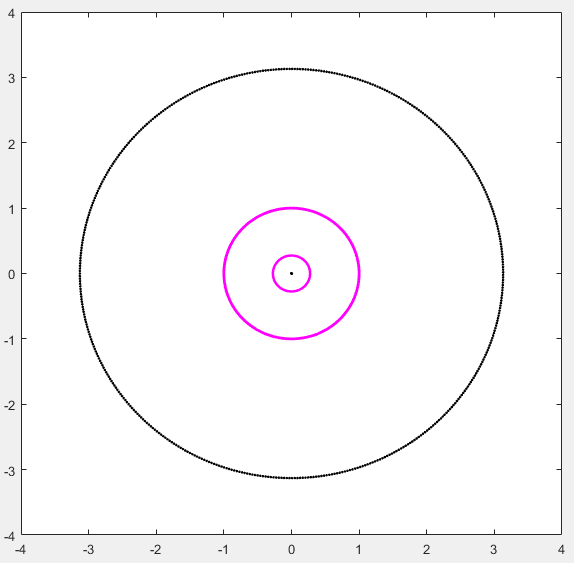}
    \includegraphics[height=5cm]{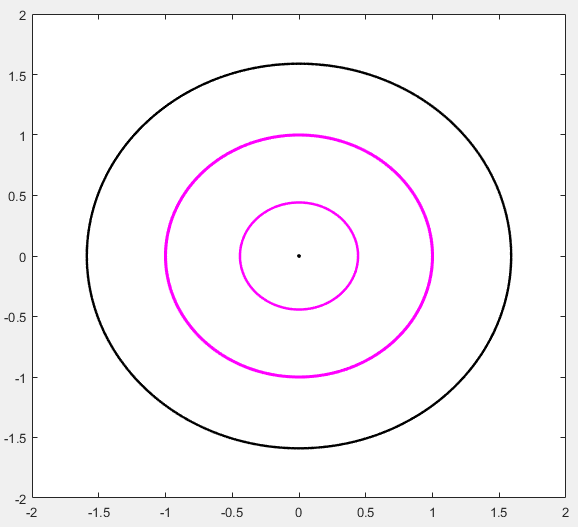}  
    \includegraphics[height=5cm]{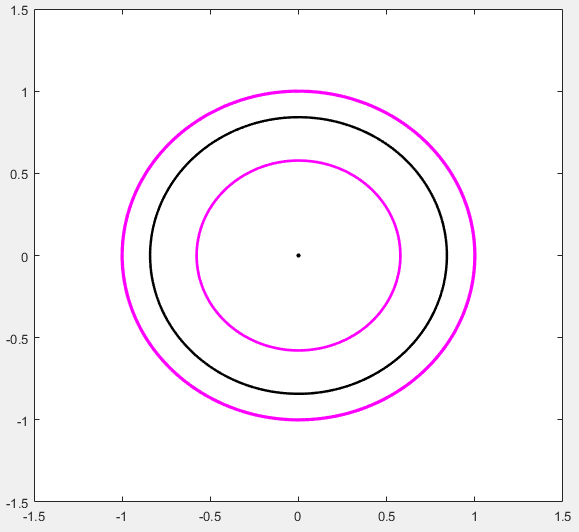}\\
    \includegraphics[height=5cm]{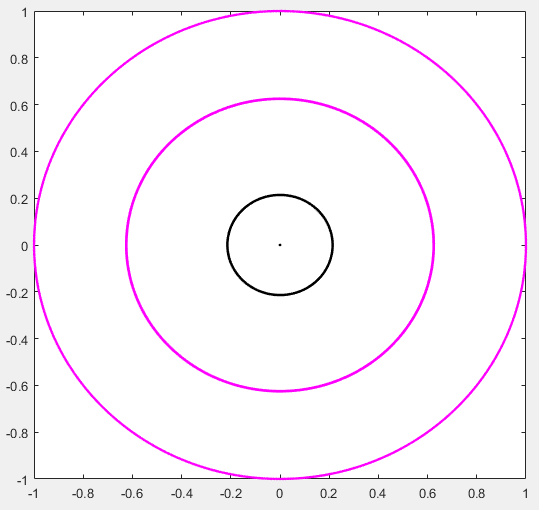} 
    \includegraphics[height=5cm]{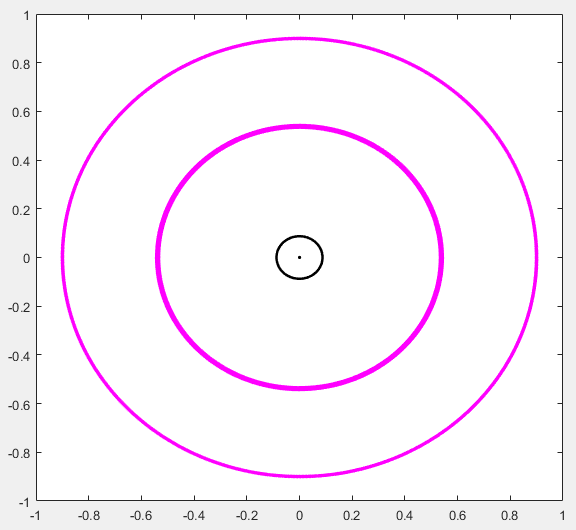} 
    \includegraphics[height=5cm]{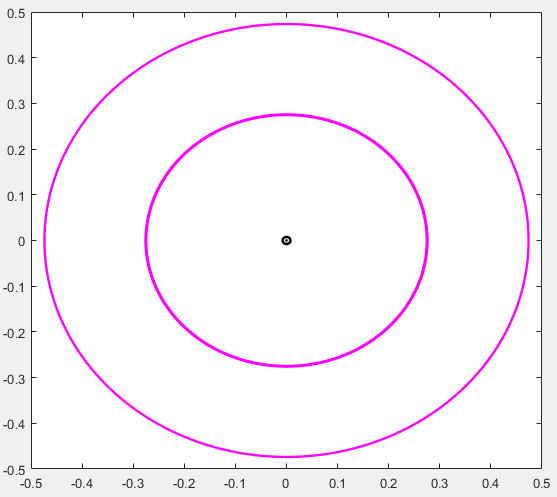} 
    \caption{Caustics and Critical Curves for the uniform transparent sphere, for $\boldsymbol{M}=\boldsymbol{I}$. The critical curves are in magenta and the caustics in black. The top row has respectively, from left to right, $x_{0}=\{0.3,0.5,0.7\}$. The bottom row has values, from left to right, of $x_{0}=\{1,1.1,1.2\}$.}
\end{figure}

\noindent
In Figure 5, there is also an odd number of images. The images stopped being connected once the source is not on the origin of the mass distribution.

\noindent
Another interesting result is that the number of images goes from five to three in Figure 5 (columns 1 and 2). This  is  yet  another  common  result  from gravitational  lensing. If  the  source  crosses  the  caustic,  the  number  of images is reduced by two \cite{SEF}. In Figure 5, the caustic crossing can be seen.  The images shown in Figure 6 are the critical curve and caustic if $x_{0}=1$. The difference here is that both the diamond and the elipse form as caustic solutions, and in the critical curves, two ellipse-like curves are forming as solutions

\begin{figure}[H]
\centering
    \includegraphics[width=.35\textwidth]{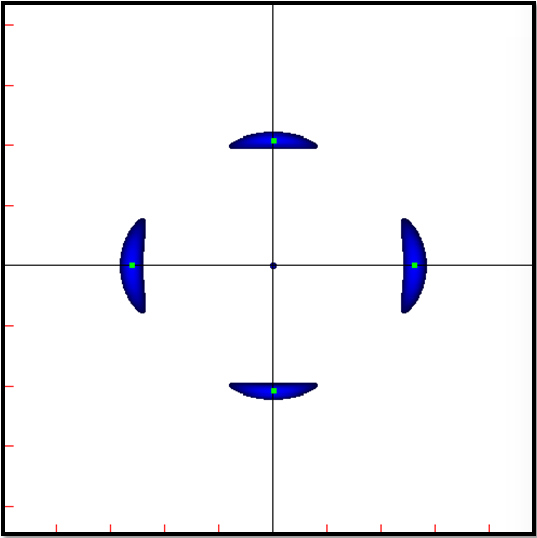} 
    \includegraphics[width=.35\textwidth]{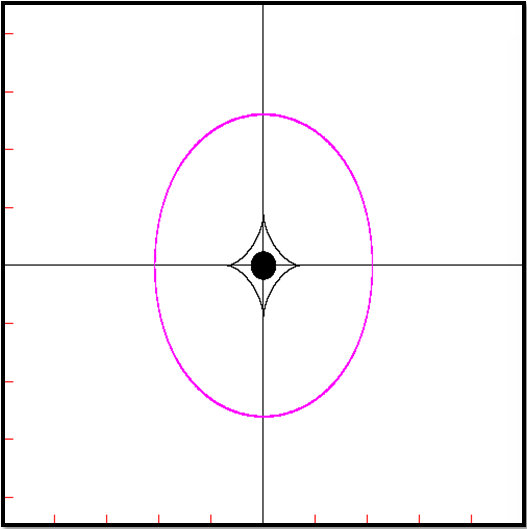}\\[\smallskipamount]
    \includegraphics[width=.35\textwidth]{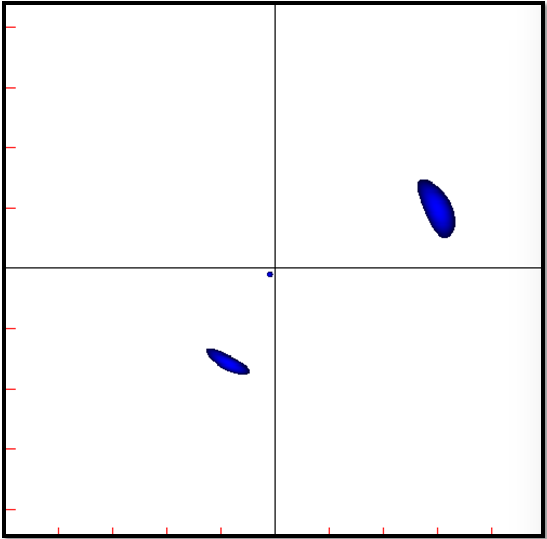}
    \includegraphics[width=.35\textwidth]{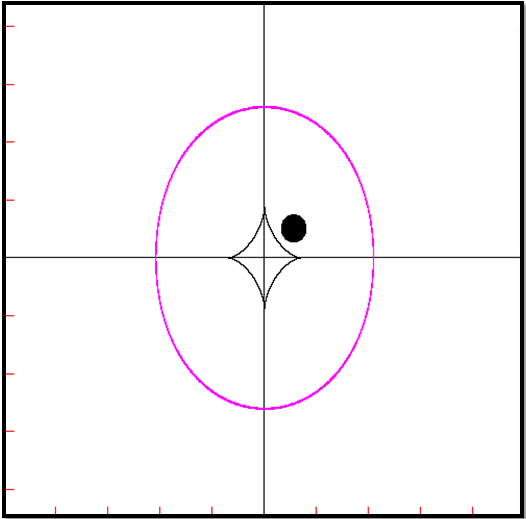}\\[\smallskipamount]
    \includegraphics[width=.35\textwidth]{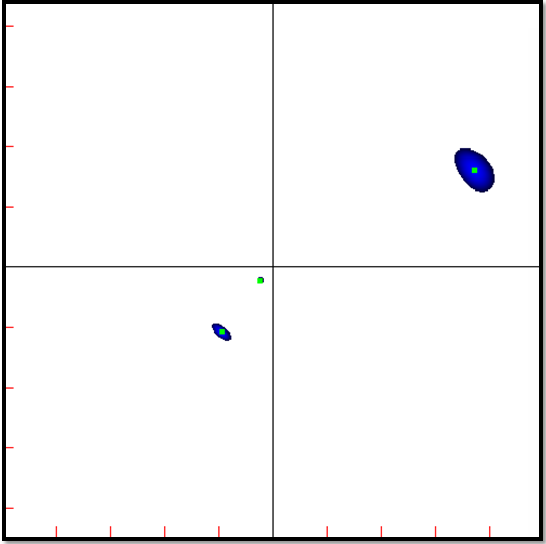}
    \includegraphics[width=.35\textwidth]{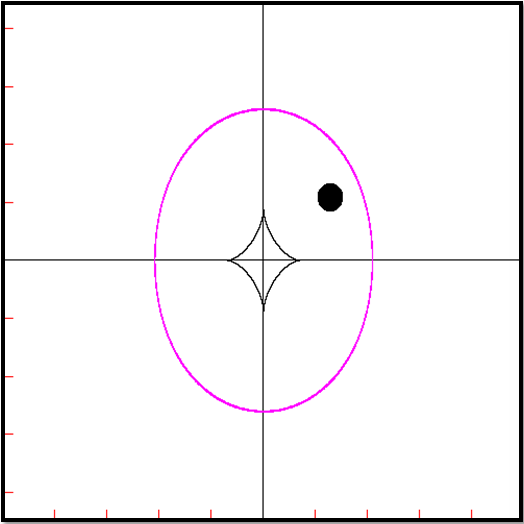}
    \caption{Images (left) and Critical Curves and Caustics (right) for the uniform transparent sphere. The black dot represents where the original source is with respect to the center of the distribution, and the green dots represent individual images. The values for the parameters are: $\sigma = 0.25$, $\phi = 0$ and $\gamma = 0.12$ and $x_{0}=1$. }\label{fig:foobar1}
\end{figure}

\begin{figure}[H]
    \centering
    \includegraphics[height=6cm]{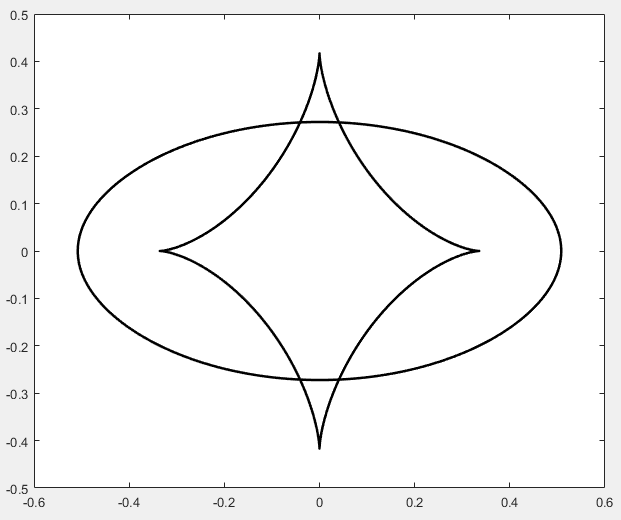}
    \includegraphics[height=6cm]{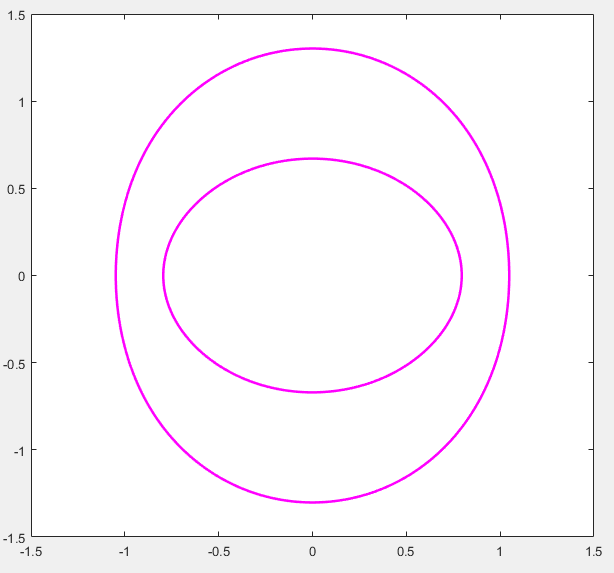}
    \caption{Images of the caustics (left) and the critical curves (right) plotted for the uniform transparent sphere. The parameters are $\phi=0$, $\sigma = 0.25$ and $\gamma = 0.16$. and $x_{0}=1$.}
\end{figure}

\begin{figure}[h]
    \centering
    \includegraphics[width=.38\textwidth]{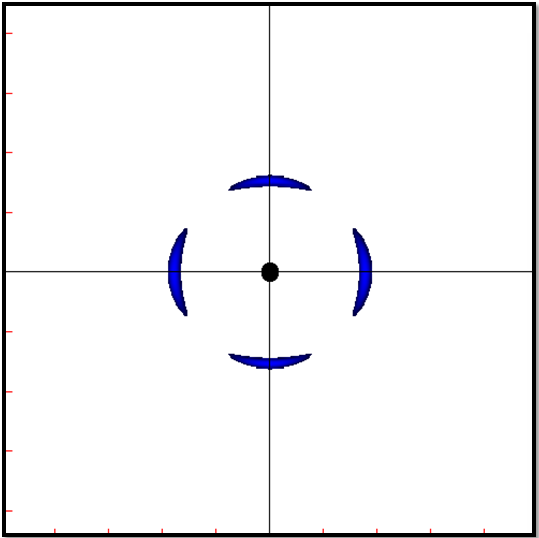}
    \includegraphics[width=.38\textwidth]{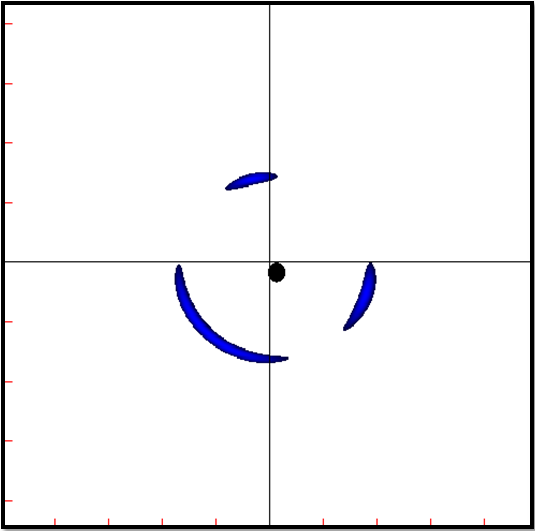}\\[\smallskipamount]
    \includegraphics[width=.38\textwidth]{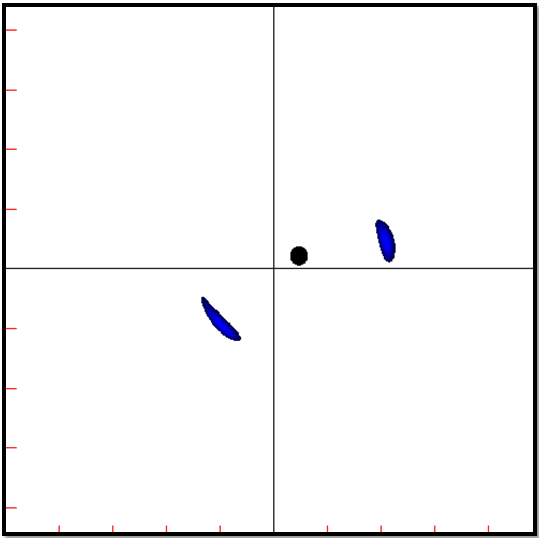}
    \includegraphics[width=.38\textwidth]{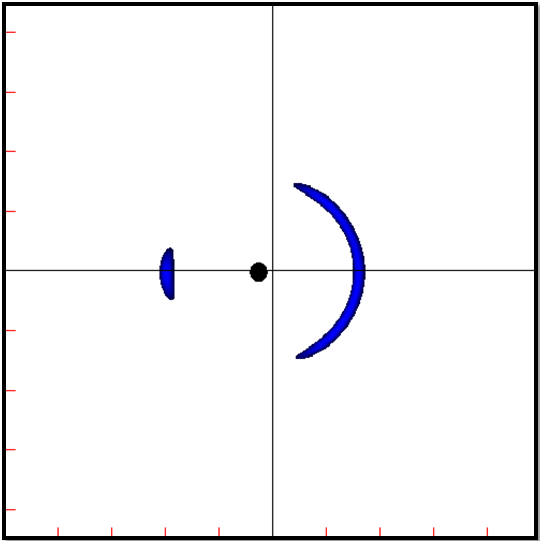}
    \caption{Lens images formed by a SIS distribution. The black dot represents where the original source is with respect to the center of the distribution. The parameters are $\phi=0$, $\sigma = 0.16$ and $\gamma = 0.12$. }\label{fig:foobar2}
\end{figure}

\newpage
\noindent
From Figure 7, we have a particularly interesting result, because the central image that appeared in the other models does not appear in this case. The reason this occurs is that the central image is the result of the unlensed light from the source going perpendicularly through the lens and to the observer. However, the SIS model has a divergence in $r=0$ that results in the equation not giving this particular solution. This is consistent with the theory that predict an odd number of images, because these articles started from the assumption that the mass distribution was physically sensible, and that it did not have a divergence at the origin. In 1980, Dyer and Roeder mentioned that they deduced the odd number of images for a transparent profile assuming that the matter distribution diverges less rapidly than $1/r$ as $r$ goes to $0$, which is not the case for the SIS profile \cite{DR}. What we observe in the images correspond to the images that come in pairs in the transparent case, and not the image that goes from the source directly to the observer.

\begin{figure}[H]
    \includegraphics[height=7.8cm]{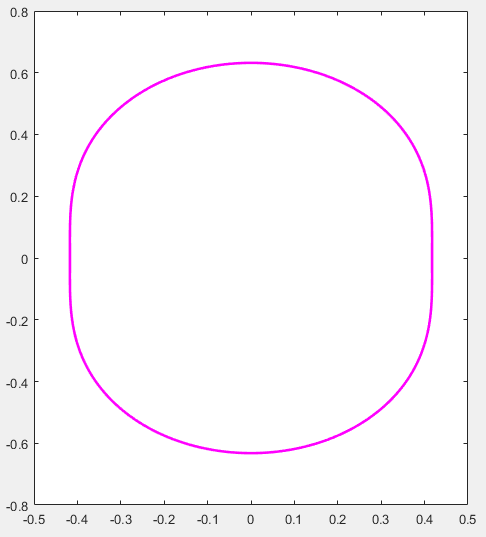}\hfill
    \includegraphics[height=7.8cm]{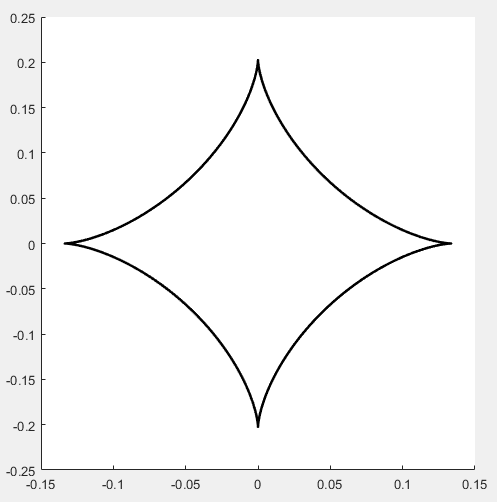}
    \caption{Images of the caustics (left) and the critical curves (right) plotted for the Transparent SIS. The parameters are $\phi=0$, $\sigma = 0.25$, $\gamma = 0.16$ and $x_{0}=1$. }
\end{figure}

\noindent
In Figure 8, it can be seen that the caustic solution for the transparent SIS profile is a diamond, and the critical curve it is an ellipse. These are very common as shapes for caustics and critical curves, even though they are not the only possible solutions.
 
\noindent
The NSIS case is presented in Figure 9. In this case, four symmetrical images instead of an Einstein ring are observed, and the image from the center that goes through the matter distribution, because the lens is transparent. This is known as an Einstein cross, and it has been observed \cite{SEF}. Five images appear in the top row and in the bottom left image, when one considers that the arc contains 3 images. In the bottom right image, the source is outside the distribution, resulting in two images from the point case.

\begin{figure}[H]
    \centering
    \includegraphics[width=.30\textwidth]{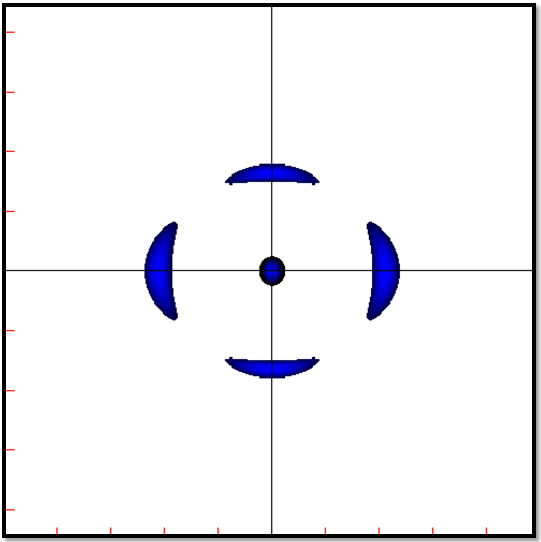}
    \includegraphics[width=.30\textwidth]{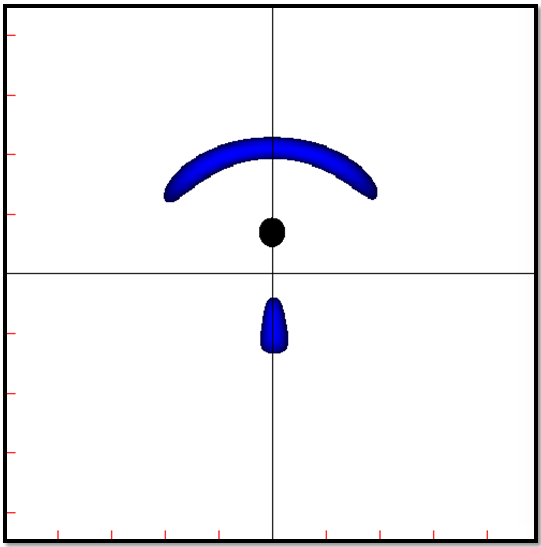}\\[\smallskipamount]
    \includegraphics[width=.30\textwidth]{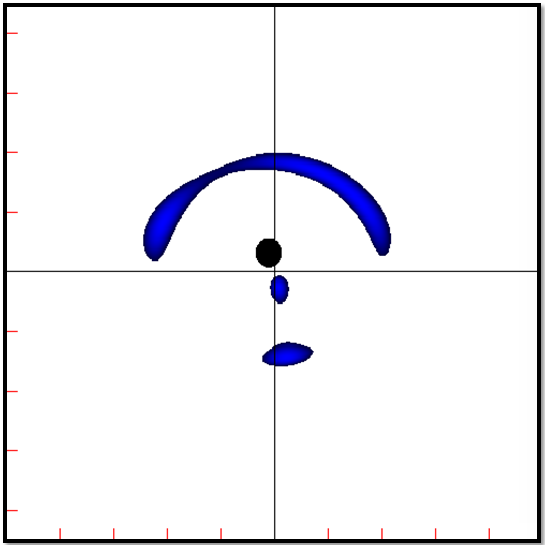}
    \includegraphics[width=.30\textwidth]{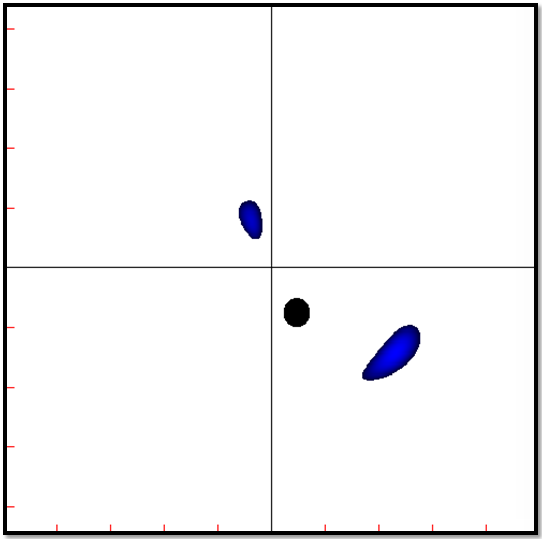}
    \caption{Lens images formed by a NSIS distribution. The black dot represents where the original source is with respect to the center of the distribution. The parameters are $\phi=0$, $\sigma = 0.33$ and $\gamma = 0.16$. }\label{fig:foobar3}
\end{figure}

\begin{figure}[H]
     \centering
    \includegraphics[height=8cm]{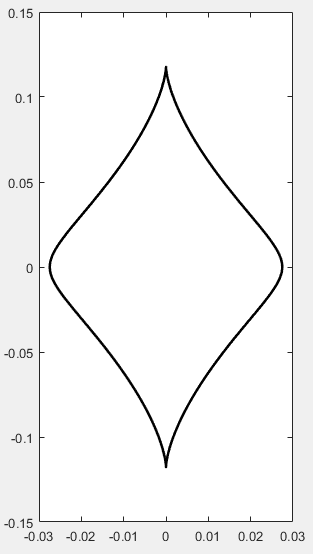}
    \includegraphics[height=8cm]{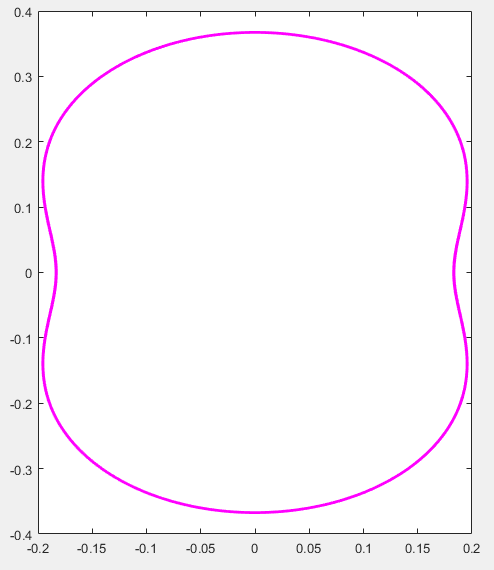}
    \caption{Images of the caustics (left) and the critical curves (right) plotted for the transparent NSIS. The parameters are $\phi=0$, $\sigma = 0.25$, $\gamma = 0.16$ and $x_{0}=1$. }
\end{figure}


\noindent
In Figure 10, the caustic and critical curves for NSIS are quite different from the singular case. The graph for the caustic has a diamond-like figure, and the plot for the critic curve has a lemniscate-like figure. 

\begin{figure}[H]
    \centering
    \includegraphics[width=.45\textwidth]{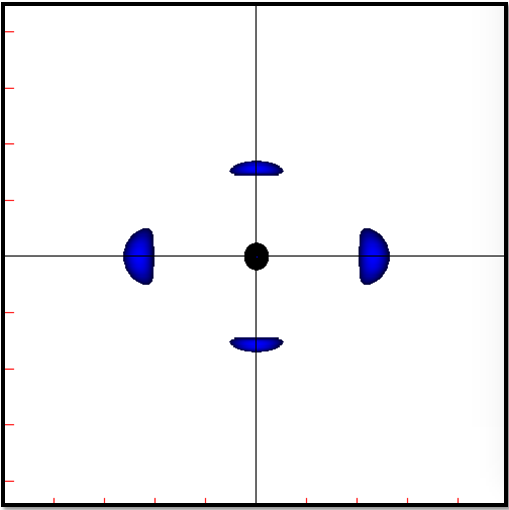}
    \includegraphics[width=.45\textwidth]{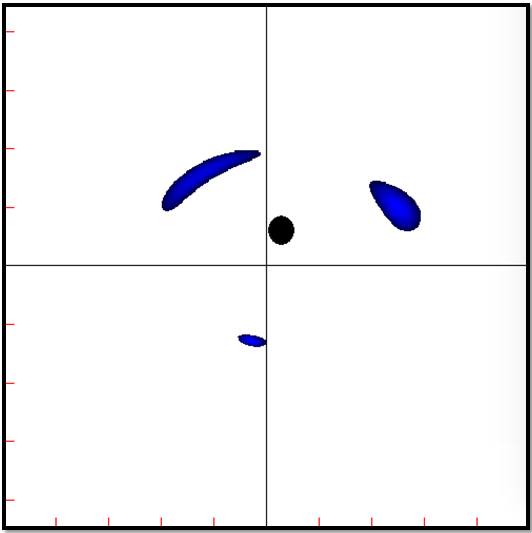}\\[\smallskipamount]
    \includegraphics[width=.45\textwidth]{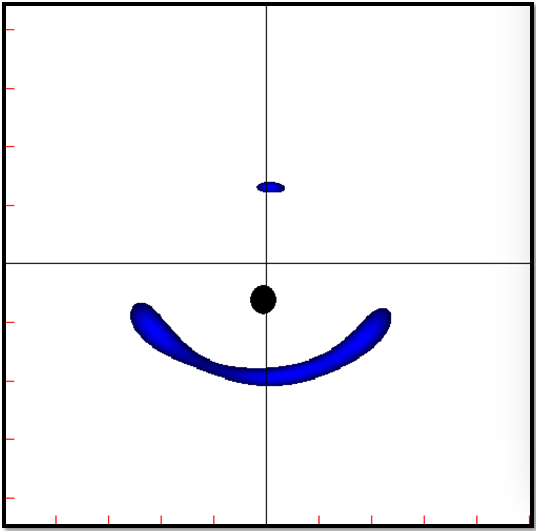}
    \includegraphics[width=.45\textwidth]{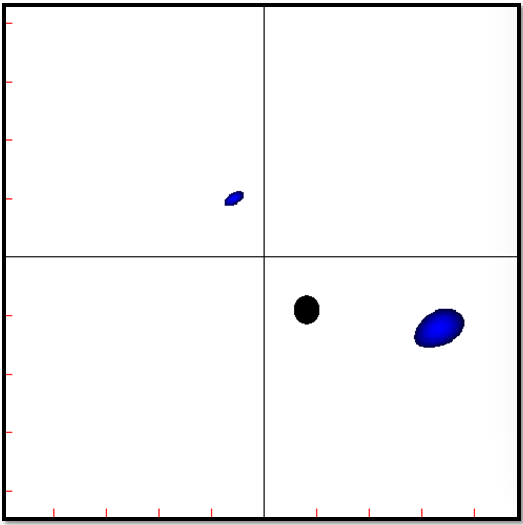}
    \caption{Lens images formed by a transparent King distribution. The black dot represents where the original source is with respect to the center of the distribution. In the 4 images, $\phi=0$. For $\gamma$, in the first row and bottom left image $\gamma = 0.16$ and in the bottom right $\gamma = 0.08$. For $\sigma$, the first column have  $\sigma = 0.33$ and the second column $\sigma = 0.25$. In the Einstein cross, the central image is little and opaqued by the source. }\label{fig:foobar4}
\end{figure}

\begin{figure}[H]
     
    \includegraphics[height=7cm]{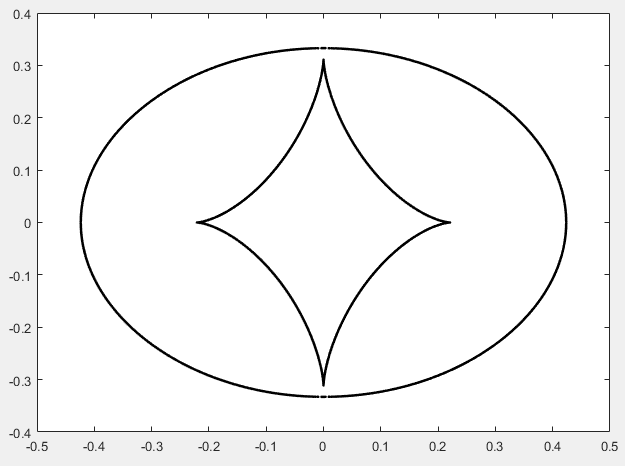}
    \includegraphics[height=7cm]{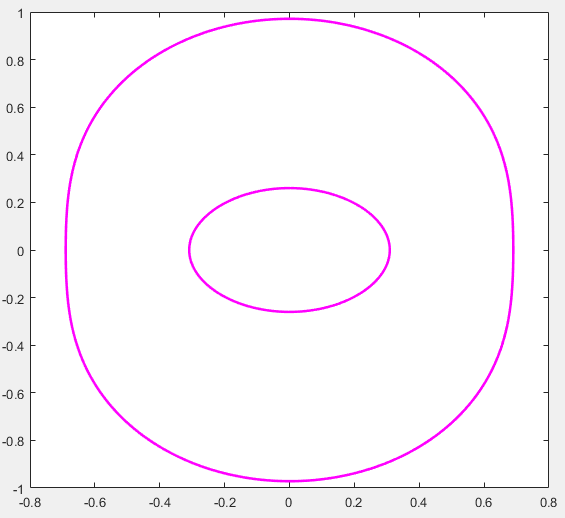}
    \caption{Images of the caustics (left) and the critical curves (right) plotted for the transparent King. The parameters are $\phi=0$, $\sigma = 0.25$, $\gamma = 0.16$ and $x_{0}=1$. }
\end{figure}

\noindent
The same interesting phenomenon as in the uniform transparent sphere is shown in Figures 11 and 12. There are two critical curves and two caustics. As in the transparent NSIS profile, if the source is at the origin, an Einstein cross is observed. Another phenomenom, we can observe from Figure 10 is that changing the values $ \gamma, \, \phi, \, \sigma $ of $ \boldsymbol{M} $, the specific shape for the Einstein cross can change, as is shown from the top left to the top right image. Different values for these parameters can be adjusted so that a real image can be mapped with a model. $\phi$ has the effect of rotating the images in this case. In Figure 10, we can also observe the transtition from five to three and then to one image. Note that in the bottom left image, the arc embodies two images.

\section{Conclusions}

In this contribution, we present an extensive explanation regarding macro-gravitional lenses and how to calculate different properties of this images in the case of a transparent distribution of matter, following a specific profile. 

\noindent
With the help of XFGLenses, and with MATLAB, we show different images that arise from all of these profiles, and the different caustics and critical curves. The images are consistent with several previous results that are expected for transparent profiles. One of them is that these give rise to an odd number of images \cite{DR, McKenzie}. The other one is that if the sources passes through the caustic, the number of images is reduced by two \cite{SEF}. Finally, the curves shown in the caustics where the diamond, the ellipse and the lemniscate-like. For the critical curves, the most common curve is the ellipse, and the lemniscate-like appear in the transparent NSIS case, which is consistent with the fact that these curves are common in gravitational lens theory.


\end{document}